\documentclass[article,preprint,groupedaddress,amsmath,amssymbm]{revtex4}
\usepackage{epsfig}
\usepackage{graphicx}

\begin{document}
\title{Ten shades of black}
%\title{Quantum black holes and large extra dimensions}
%\title{Can we see the large extra dimensions?}
\author{Shahar Hod}
\address{The Ruppin Academic Center, Emeq Hefer 40250, Israel}
\address{ }
\address{The Hadassah Institute, Jerusalem 91010, Israel}
\date{\today}
%{\it Essay written for the Gravity Research Foundation 2015 Awards
%for Essays on Gravitation}
{\it This essay received an Honorable Mention in the Gravity
Research Foundation Essay Competition 2015.}

\begin{abstract}

\ \ \ The holographic principle has taught us that, as far as their
entropy content is concerned, black holes in $(3+1)$-dimensional
curved spacetimes behave as ordinary thermodynamic systems in flat
$(2+1)$-dimensional spacetimes. In this essay we point out that the
{\it opposite} behavior can also be observed in black-hole physics.
To show this we study the quantum Hawking evaporation of
near-extremal Reissner-Nordstr\"om black holes. We first point out
that the black-hole radiation spectrum departs from the familiar
radiation spectrum of genuine $(3+1)$-dimensional perfect black-body
emitters. In particular, the would be black-body thermal spectrum is
distorted by the curvature potential which surrounds the black hole
and effectively blocks the emission of low-energy quanta. Taking
into account the energy-dependent gray-body factors which quantify
the imprint of passage of the emitted radiation quanta through the
black-hole curvature potential, we reveal that the
$(3+1)$-dimensional black holes effectively behave as perfect
black-body emitters in a flat $(9+1)$-dimensional spacetime.
\newline
\newline
Email: shaharhod@gmail.com
\end{abstract}
\bigskip
\maketitle

%]

{\bf Introduction. --- } The holographic principle has revealed
that, as far as their thermodynamic properties are concerned, black
holes in $(3+1)$-dimensional spacetimes are fundamentally
$(2+1)$-dimensional objects \cite{Hof,Sus,BekMay}. In particular,
the entropy content of a black hole in $3$-D space scales with the
black-hole $2$-D surface area \cite{Bek}. In this essay we reveal
that the {\it opposite} behavior can also be observed in the physics
of black holes.

To that end, we study the quantum Hawking evaporation of
near-extremal Reissner-Nordstr\"om (RN) black holes. We shall show
that the effective curvature potential which surrounds the
evaporating black holes distorts the emitted Hawking quanta in such
a way that the resulting black-hole radiation spectrum is no longer
that of a perfect 3-D black-body emitter. Moreover, a detailed
analysis (to be carried out below) reveals that these
$(3+1)$-dimensional black holes effectively behave as perfect
black-body emitters in a flat $(9+1)$-dimensional spacetime.

{\bf Quantum evaporation of near-extremal RN black holes. --- }
Hawking's celebrated result that black holes emit a thermally
distributed radiation is certainly one of the most important
theoretical predictions of modern physics \cite{Haw}. Although the
Hawking black-hole radiation spectrum has distinct thermal features,
it is important to realize that it departs from the familiar
radiation spectra of perfect black-body emitters. In particular, the
would be black-body thermal spectrum is distorted by the curvature
potential which surrounds the black hole and effectively blocks the
emission of low-energy quanta. The departure of the Hawking
black-hole radiation spectrum from the pure spectrum of a perfect
(thermal) black-body emitter can be quantified by the
frequency-dependent gray-body factors $\{\Gamma(\omega)\}$
\cite{Page}.

In this essay we shall explore the power spectrum of the coupled
electromagnetic-gravitational quanta emitted by near-extremal
Reissner-Nordstr\"om (RN) black holes. The Hawking temperature of
these $(3+1)$-dimensional black holes is given by \cite{Noteunit}
\begin{equation}\label{Eq1}
T_{\text{RN}}={{\hbar\sqrt{M^2-Q^2}}\over{2\pi(M+\sqrt{M^2-Q^2})^2}}\
,
\end{equation}
where $M$ and $Q$ are the black-hole mass and electric charge,
respectively. The Hawking radiation power out of the black holes is
given by \cite{Zurek}
\begin{equation}\label{Eq2}
P^{3+1}_{\text{RN}}={{T^{\text{RN}}}\over{2\pi}}\sum_{l,m}\int_0^{\infty}
d\omega{{\Gamma x}\over{e^x-1}}\  ,
\end{equation}
where $x\equiv\hbar\omega/T_{\text{RN}}$. Here $l$ and $-l\leq m\leq
l$ are the harmonic indexes of the emitted quanta, and
$\Gamma=\Gamma_{lm}(\omega)$ are the energy-dependent gray-body
factors \cite{Page}. These dimensionless transmission coefficients
quantify the imprint of passage of the emitted radiation through the
effective curvature potential which surrounds the black hole.

The thermal factor that appears in the denominator of (\ref{Eq2})
implies that the black-hole emission spectrum peaks at the
characteristic frequency $x^{\text{peak}}\equiv
\hbar\omega^{\text{peak}}/T_{\text{RN}}=O(1)$. Remembering that
near-extremal black holes are characterized by the relation
$MT_{\text{RN}}/\hbar\ll1$, one finds the strong inequality
\begin{equation}\label{Eq3}
M\omega^{\text{peak}}\ll1\
\end{equation}
for the characteristic frequencies emitted by the near-extremal
black holes.

The relation (\ref{Eq3}) implies that, for near-extremal black
holes, the typical wavelengths in the Hawking radiation spectrum are
very large on the scale set by the geometric size of the evaporating
black hole. The calculation of the frequency-dependent grey-body
factors $\Gamma_{lm}(\omega)$ in the low-frequency regime
(\ref{Eq3}) is a common practice in the physics of black holes
\cite{Page}. In particular, one finds the leading-order behavior
\cite{Cris}
\begin{equation}\label{Eq4}
\Gamma_{11}=\Gamma_{2m}={4\over9}(\omega r_{\text{H}})^8\
\end{equation}
in the small-frequency regime $\omega r_{\text{H}}\ll1$, where
$r_{\text{H}}\simeq M$ is the outer horizon radius of the
near-extremal black hole \cite{Notell}. Substituting the
frequency-dependent grey-body factors (\ref{Eq4}) into (\ref{Eq2})
and performing the integration, one finds
\begin{equation}\label{Eq5}
P^{3+1}_{\text{RN}}=C^{3+1}_{\text{RN}}\times{{r^8_{\text{H}}{T^{10}_{\text{RN}}}}\over{\hbar^9}}\
\end{equation}
for the emission power out of the RN black holes \cite{Notern}.

Evidently, the Hawking radiation power (\ref{Eq5}) out of the
$(3+1)$-dimensional near-extremal black holes looks completely
different from the familiar Stefan-Boltzmann law \cite{Allen}
\begin{equation}\label{Eq6}
P^{3+1}_{\text{flat}}=C^{3+1}_{\text{flat}}\times{{R^2T^4}\over{\hbar^3}}\
\end{equation}
for perfect black-body emitters of temperature $T$ and radius $R$ in
a $(3+1)$-dimensional flat spacetime \cite{Notecf}.

We shall now prove, however, that the Hawking radiation power
(\ref{Eq5}) characterizing the $(3+1)$-dimensional near-extremal
black holes is of the {\it same} functional form as the thermal
radiation power of perfect black-body emitters in a {\it
higher}-dimensional flat spacetime.

{\bf Perfect black-body emitters in flat $(D+1)$-dimensional
spacetimes. --- } We shall now obtain the thermal radiation power
which characterizes perfect black-body emitters in general
$(D+1)$-dimensional flat spacetimes.

To that end, we first note that the thermal energy density of one
bosonic degree of freedom inside a $(D+1)$-dimensional closed cavity
of temperature $T$ is given by \cite{BekHod,Carn}
\begin{equation}\label{Eq7}
\rho_D={{T}\over{{(2\pi)}^D}}{\int_0^{\infty}}\
dV_D(\omega){{x}\over{{e^x-1}}}\ ,
\end{equation}
where $x\equiv \hbar\omega/T$ and
%\begin{equation}\label{Eq4}
$dV_D(\omega)=[2\pi^{D/2}/\Gamma(D/2)]\omega^{D-1}d\omega$
%\end{equation}
is the volume in the $(D+1)$-dimensional frequency-space of the
shell $(\omega,\omega+d\omega)$. Substituting $dV_D(\omega)$ into
(\ref{Eq7}) and performing the integration, one finds the
$(D+1)$-dimensional thermal energy density
\begin{equation}\label{Eq8}
\rho_D={{\Gamma(D+1)\zeta(D+1)}\over{2^{D-1}\pi^{D/2}\Gamma(D/2)}}\times
{{T^{D+1}}\over{\hbar^D}}\  ,
\end{equation}
where $\zeta(z)$ is the Riemann zeta function \cite{Notezet}. Since
the thermal radiation is emitted from a sphere of surface-area
$A_{D-1}=[{{2\pi^{D/2}}/{\Gamma(D/2)}}]R^{D-1}$ \cite{Notesp}, the
radiated power $P^{D+1}_{\text{flat}}$ out of the
$(D+1)$-dimensional perfect black-body emitter is proportional to
$\rho_D\times A_{D-1}$, which yields \cite{Carn,Notecd}:
\begin{equation}\label{Eq9}
P^{D+1}_{\text{flat}}=C^{D+1}_{\text{flat}}\times
{{R^{D-1}T^{D+1}}\over{\hbar^D}}\  .
\end{equation}

A direct comparison between the two radiation powers, Eqs.
(\ref{Eq5}) and (\ref{Eq9}), reveals the surprising conclusion that
our $(3+1)$-dimensional near-extremal black holes effectively behave
as perfect black-body emitters in a flat $(9+1)$-dimensional
spacetime.

{\bf Summary. --- } In this essay we have analyzed the Hawking
emission of coupled electromagnetic-gravitational quanta by
near-extremal Reissner-Nordstr\"om black holes. It was pointed out
that, due to the influence of the energy-dependent gray-body factors
which quantify the imprint of passage of the emitted quanta through
the spacetime curvature potential, the black-hole radiation spectrum
departs from the familiar radiation spectrum of genuine
$(3+1)$-dimensional perfect black-body emitters. In particular, it
was shown that the curvature potential which surrounds these
$(3+1)$-dimensional near-extremal black holes distorts the emitted
Hawking spectrum in such a way that the resulting black-hole
radiation power is effectively that of a perfect black-body emitter
in a flat $(9+1)$-dimensional spacetime.

\bigskip
%\newline

{\it Acknowledgments:} This research is supported by the Carmel
Science Foundation. I thank Yael Oren, Arbel M. Ongo, Ayelet B.
Lata, and Alona B. Tea for stimulating discussions.


\begin{thebibliography}{99}
\bibitem{Noteunit} We use gravitational units in which $G=c=k_{\text{B}}=1$.
\bibitem{Notell} Note that the emission of quanta with larger values of the
spherical harmonic index $l$ is suppressed due to the fact that the
corresponding grey-body factors of these higher-$l$ modes contain
higher powers of the small quantity $\omega r_{\text{H}}\ll1$
\cite{Cris}.
\bibitem{Notern} The dimensionless coefficient in (\ref{Eq5}) is given by
$C^{3+1}_{\text{RN}}={{1024\pi^9}/{297}}$.
\bibitem{Notecf} The dimensionless coefficient in (\ref{Eq6}) is given by
$C^{3+1}_{\text{flat}}={{\pi^3}/{30}}$ \cite{Allen}.
\bibitem{Notezet} Here we have used the relation
${\int_0^{\infty}}{{dx x^D}/{(e^x-1)}}=\Gamma(D+1)\zeta(D+1)$.
\bibitem{Notesp} Note that the radiating surface-area is of $D-1$ spatial
dimensions.
\bibitem{Notecd} The dimensionless coefficient in (\ref{Eq9}) is given by
$C^{D+1}_{\text{flat}}=D\zeta(D+1)/\pi$. Here we have used the
identity $\Gamma(2z)=\pi^{-1/2}2^{2z-1}\Gamma(z)\Gamma(z+1/2)$.
\end{thebibliography}
\end{document}